\documentclass[11pt]{article}
\usepackage{amsmath}
\usepackage{amsfonts}
 \newcounter{mnotecount}[section]

 \renewcommand{\themnotecount}{\thesection.\arabic{mnotecount}}

 \newcommand{\mnote}[1]%{}%
 {\protect{\stepcounter{mnotecount}}$^{\mbox{\footnotesize
 $%\!\!\!\!\!\!\,
 \bullet$\themnotecount}}$ \marginpar{%\color{red}%
 \raggedright\tiny\em
 $\!\!\!\!\!\!\,\bullet$\themnotecount: #1} }

\def\be{\begin{equation}}
\def\ee{\end{equation}}
\def\bea{\begin{eqnarray}}
\def\eea{\end{eqnarray}}

\def\hR{\hat{R}}
\def\hg{\hat{g}}
\def\hM{\hat{M}}
\def\hOmega{\hat{\Omega}}
\def\htt{\hat{t}}

\def\scri{\mathcal{I}}
\def\hM{\hat{M}}
\def\hrho{\hat{\rho}}
\def\hs{\hat{s}}
\def\hG{\hat{G}}

\def\hg{\hat{g}}
\def\hM{\hat{M}}

\def\hOmega{\hat{\Omega}}

\def\cR{\check{R}}

\def\cg{\check{g}}

\def\cM{\check{M}}

\def\ct{\check{t}}

\def\cG{\check{G}}

\def\cOmega{\check{\Omega}}

\def\crho{\check{\rho}}

\begin{document}

\title{Some questions about Conformal Cyclic Cosmology}
\author{Paul Tod}

\maketitle
%%%%%%%%%%%%%%%%%%%%%%%%%%%%%%%%%%%%%%%%%
\begin{abstract}This article is an extended version of a talk given in Oxford in June 2021 as part of an online meeting `Ninety minutes of CCC' to mark the 90th birthday of Sir Roger Penrose. I assemble some questions that I have been asked or have asked myself about CCC.

\end{abstract}
%%%%%%%%%%%%%%%%%%%%%%%%%%%%%%%%%%%%%%%%%%
%PACS numbers:

\section*{Introduction and sketch of Conformal Cyclic Cosmology}
In this article I collect together some questions about Penrose's {\it{Conformal Cyclic Cosmology}} \cite{rp1}  (hereafter CCC) that I've been asked or have asked myself, particularly when giving the lectures \cite{t3}. My point of view is that CCC needs to answer these questions but I see no reason to think it can't.

I'll begin with a sketch of CCC. Assuming the presence in the universe of a positive cosmological constant $\Lambda$ there is a firm consensus that the universe is expanding away from a hot big bang with infinite density, through an era of galaxies into a period of exponential expansion. We model the universe as a solution of a suitable Einstein equation and then, with positive $\Lambda$,  Penrose's conformal methods indicate that there will be a conformal rescaling of the space-time metric in the remote future which can add a future infinity $\scri^+$ as a space-like boundary. It is natural to suppose that the rescaled metric is smooth there -- natural in the sense that one can construct many examples with that property, and in fact, following \cite{f1}, one can give well-posed Cauchy problems with data at a space-like $\scri$ for many matter models. 

The cosmological model will have a curvature singularity at the beginning, and Penrose has long argued that, as a singularity of a Lorentzian manifold, this big bang must be rather special, much less general than such singularities can be. Penrose's \emph{Weyl Curvature Hypothesis} \cite{rp3} is the hypothesis that the Weyl curvature is zero, or at least finite, at any initial singularity and therefore at the big bang. It is tricky, though possible (see e.g. \cite{lt}), to give a statement of what it means to have `finite Weyl tensor' at a singularity of the Riemann tensor (for example, the singularity is not yet a `place' where things can happen) and a simpler condition to impose is that there should be a conformal rescaling of the space-time metric which removes the singularity and adds an initial space-like boundary hypersurface. This is a strong condition, apparently stronger than the bare Weyl Curvature Hypothesis, but it will follow from a strong enough version of the Weyl Curvature Hypothesis and conditions of uniformity, \cite{lt}.

Now comes Penrose's `outrageous suggestion': the Weyl Curvature Hypothesis, which is motivated by observation as Penrose explains \cite{rp3}, takes it simplest form if the space-time metric can be scaled \emph{up} to allow a past space-like boundary to be added; the presence of positive $\Lambda$ implies that the space-time metric can be scaled \emph{down} to allow a future space-like boundary to be added; we have an {\it{aeon}} of universe between initial and final space-like surfaces through both of which the conformally-related metric can be extended; so let us hypothesise a continuation of the conformal metric in both directions to a preceding aeon and a subsequent aeon; so there would be a single conformal metric but different physical metrics in each aeon, all of them conformal to the single conformal metric. At the future boundary of an aeon, the Weyl curvature necessarily vanishes (though there are free degrees of freedom in its normal derivative) so with this identification the Weyl tensor will also necessarily vanish at the big bang of any aeon -- by this means the Weyl Curvature Hypothesis becomes a deduction.

To introduce equations we consider two successive aeons, say the previous one, characterised by hatted quantities, and the present one characterised by quantities with a check. Then the previous aeon is a space-time $\hM$ with metric $\hg$ and the present is $\cM$ with $\cg$. There is a conformal metric $g$ defined right through, including the surface $\scri$ which is the future boundary of $\hM$ and the past boundary of $\cM$. Strictly speaking $\scri$ is in neither $\hM$ nor $\cM$ but we assume as much regularity on the manifold $M:=\hM\cup\cM\cup\scri$ as we need. Now we have conformal factors $\hOmega,\cOmega$ such that
\be\label{1}\hg=\hOmega^2g,\;\;\cg=\cOmega^2g,\ee
with
\[\hOmega\rightarrow\infty,\;\;\cOmega\rightarrow 0 \mbox{  towards }\scri.\]
We make another regularity assumption, that the product $\hOmega\cOmega$ is finite and nonzero in a neighbourhood of $\scri$. The product is negative if we suppose that each of $\hOmega,\cOmega$ is positive in its own $\hM,\cM$. At this stage there is conformal freedom in $g$: we can make the changes
\[(\hOmega,\cOmega, g)\rightarrow(\theta\hOmega,\theta\cOmega,\theta^{-2}g),\]for positive smooth $\theta$. This is a gauge freedom -- the two physical metrics $\hg,\cg$ are unchanged by it -- which evidently changes the product of the conformal factors
\[\hOmega\cOmega\rightarrow\theta^2\hOmega\cOmega,\]
so that we can choose $\theta$ to set
\be\label{2}\hOmega\cOmega=-1.\ee
This is Penrose's {\it{Reciprocal Hypothesis}} \cite{rp1} which I'm here regarding as a gauge condition. Given (\ref{2}), we have $\cOmega$ in terms of $\hOmega$ and therefore
\be\label{2b}\cg=\cOmega^2g=\cOmega^2\hOmega^{-2}\hg=\hOmega^{-4}\hg,\ee
so that the physical metric in the present aeon is determined by the physical metric in the previous aeon and $\hOmega$. This sets the scene for question 1 below.

\medskip

It may be helpful at this point to recall a concrete example from \cite{t3}. Consider the FLRW metric
\be\label{2a}ds^2=dt^2-R(t)^2d\sigma_k^2\ee
where $d\sigma_k^2$ is one of the standard constant-curvature 3-metrics. Take the matter content to be radiation fluid plus the cosmological constant $\Lambda$. The Einstein field equations reduce to the conservation equation, which can be integrated to give the energy density of the radiation as
\[\rho=\mu/R^4,\]
with a constant of integration $\mu$, together with the Friedmann equation, which we'll write in terms of {\it{conformal time}}. This is defined by
\[d\tau=dt/R,\]
(remember: the great virtue of conformal time is that future light-cones of the origin in FLRW have equations $\tau-r=$ constant). Then the Friedmann equation is\footnote{To avoid the possiblity of a recollapsing universe we'll assume that $\mu\Lambda>3/4$ if $k=1$.}
\be\label{3}\left(\frac{dR}{d\tau}\right)^2=\mu -kR^2+\frac13\Lambda R^4.\ee
This equation has a striking symmetry: set $S=c_1R^{-1}$ where $c_1=(3\mu/\Lambda)^{1/2}$, then calculate
\[\left(\frac{dS}{d\tau}\right)^2=\mu -kS^2+\frac13\Lambda S^4,\]
which is precisely the Friedmann equation back again. It's worth noting that $R^{-1}$ is integrable at both ends (i.e. large and small positive $t$) so that there is only a finite amount of conformal time to cover the infinite range in proper time from big bang to $\scri$. If we set $\tau=0$ at the initial singularity, where conventionally $t=0$, then at $\scri$
\[\tau=\tau_F=\int_0^\infty R^{-1}dt,\]
writing $\tau_F$ for the value of $\tau$ at $\scri$. Since for large $t$ we have $R\sim e^{Ht}$ where $\Lambda=3H^2$ we'll also have, for large $t$
\be\label{2c}\tau_F-\tau=\int_t^\infty R^{-1}dt\sim H^{-1}e^{-Ht}\ee
which gives a measure of how the $\tau$-coordinate compresses the range of the $t$-coordinate towards $\scri$. The same point is made by noting that
\[R\sim\frac{1}{H(\tau_F-\tau)} \mbox{  towards  }\scri,\]
so that $R$ has a simple pole at $\tau=\tau_F$.

For use in Section 5, we'll put some numbers into these expressions. Following a calculation in \cite{t3}, one could use an expression for $R(t)$ for the actual current universe, including terms for the known dust and radiation densities and the measured $\Lambda$ in the Friedmann equation, to calculate that at the present epoch, at a time $1.4\times 10^{10}$ years after the bang, the ratio $\tau/\tau_F$ is about .74 -- in conformal time we are that close to the end! If we make $t$ ten times bigger, so $10^{11}$ years, then the fractional amount of conformal time remaining drops from .26 to $10^{-4}$. The whole infinite future of the universe in $t$-time is compressed to very little in $\tau$-time.

\medskip

To connect the FLRW metric to CCC, suppose the metric $\hg$ in the previous aeon takes the form of (\ref{2a}), so
\[\hg=d\htt^2-\hR(\htt)^2d\sigma_k^2=\hR^2(d\tau^2-d\sigma_k^2)\]
where we denote proper time in $\hM$ as $\htt$, and with density $\hrho=m/\hR^4$. Choose $\hOmega=c_1^{-1/2}\hR$ with $c_1$ as above, then (\ref{2b}) gives
\[\cg=\hOmega^{-4}\hg=\cR^2(d\tau^2-d\sigma_k^2)=d\ct^2-\cR^2(\ct)d\sigma_k^2,\]
with $\cR=c_1\hR^{-1}$, which by the symmetry noted therefore solves (\ref{3}). Thus the CCC prescription yields an FLRW metric in the present aeon as well, and the Einstein equations are satisfied with the `same' density, i.e. $\crho=m/\cR^4$, and same cosmological constant. In this rather special example, in which the Weyl tensor is zero throughout, all aeons are diffeomorphic.

\section{How do you define a unique $\hat{\Omega}$?}
For CCC to be well-defined, we need an algorithm for fixing the geometry in the present aeon from that in the previous aeon. By consideration of (\ref{2b}) that means we need an algorithm for a unique $\hOmega$ given $\hg$, since these then fix $\cg$. What is this algorithm to be? Penrose \cite{rp1} considers the rule for transforming scalar curvature under conformal rescaling: if metrics $g,\hg$ are related by $\hg_{ab}=\varphi^2g_{ab}$, so that $\varphi=\hOmega^{-1}$ in our previous convention, and is zero and regular at $\scri$, then the scalar curvatures -- call these $s,\hs$ respectively, to avoid confusion with the FLRW scale factor -- are related by
\be\label{2d}\widehat{\Box}\varphi+\hs\varphi=\frac16s\varphi^3,\ee
where $\widehat{\Box}$ is the d'Alembertian w.r.t. $\hg$. This equation is discussed in \cite{t3} (and elsewhere of course, starting with \cite{rp1}): if one has a Starobinsky expansion of the metric $\hg$, which is to say an expansion in powers of $e^{Ht}$ and which is not a strong assumption, then solutions of (\ref{2d}) can be expanded in negative powers as
\[\varphi=\Sigma_{n\geq 1}\varphi_n(x^i)e^{-nHt},\]
where $x^i$ are comoving spatial coordinates. Furthermore, $\varphi_1,\varphi_2$ are freely specifiable data determining all the other coefficients. Penrose imposes $\varphi_2=0$ and for reasons explained in \cite{rp1} and \cite{t3}, he calls this the {\it{delayed rest mass hypothesis}}. Then he gives a range of possibilities for fixing $\varphi_1$ without settling on one of them. In \cite{t3} I suggested a different possibility which is to choose $\varphi_1$ so that the scalar curvature of the metric induced on $\scri$ by $g$ is constant, which is what one did in the FLRW example. This is equivalent to solving the Yamabe problem for $\scri$ and in favourable circumstances this has a unique solution (but not always!).

Another approach to defining a unique $\hOmega$ (or $\varphi=\hOmega^{-1}$ which is a better behaved function in $\hM$) would be to suppose that there is a conformal scalar field, say $\phi$, in $M$ vanishing at $\scri$. Then one would choose the conformal scale to give $\phi$ the value one, which could be done throughout $\hM$ but not at $\scri$. Then one needs to decide whether $\phi$ is a physical field contributing to the Einstein equation via an enegy-momentum tensor. Something quite close to this was suggested in \cite{L} but it's not yet entirely satisfactory.

\medskip

However it is accomplished, this is a question that needs answering for CCC.

\section{What is the field content of CCC classically?}
In Penrose's original version, \cite{rp1}, the matter content of the previous aeon close to $\scri$ is assumed to be a radiation fluid, possibly with massless fields like Maxwell, and certainly with a positive cosmological constant. Earlier on, in the era of galaxies, there would be other contributions to the matter content, like a pressure-free perfect fluid, but it was important for Penrose that these dropped out near $\scri$. Without some explicit mechanism to make the dust contribution fade away, its density is proportional (in the FLRW case) to $R^{-3}$ so it comes to dominate the radiation which has density proportional to $R^{-4}$.

It follows from (\ref{2b}) how to express the Einstein tensor, say $\cG_{ab}$ in $\cM$, in terms of the Einstein tensor $\hG_{ab}$ of $\hM$ and derivatives of $\hOmega$ up to second order. Thus one at once knows the Einstein tensor and therefore the energy-momentum tensor and matter content of $\cM$ from $\hM$. This expression can be found in \cite{rp1} and \cite{t3} but it's complicated and hard to decode. In particular, it won't be anything as simple as it was in $\hM$. However it must express the matter content of $\cM$ near the big bang and therefore needs to be understood. In \cite{rp1}, Penrose proposes to understand the field $\hOmega$, or a function of it, as a conformal scalar field in the present aeon, and specifically he wants to connect this field to dark matter. As an alternative,  I've suggested that for reasons of naturalness, if there is a conformal scalar in the present aeon then perhaps there should be one in the previous aeon too, in fact in all aeons. This would connect with L\"ubbe's suggestion in \cite{L} mentioned at the end of the previous section.

\medskip

It's worth emphasising that until one has a settled view on the field content of CCC classically, one doesn't have any certainty about what the Einstein equations should be in $\cM$.

\section{What is the field content of CCC as a QFT? }
This is an area in which I have little expertise but Penrose has made a number of suggestions that are worth drawing together. The most radical is that he speculates that the rest-mass of all particles goes to zero on the approach to $\scri$, as something like the Higgs mechanism, which gives mass to the particles in the early universe, turning the mass off again. After this {\it{mass fade-out}} the matter content is massless, and one has a radiation stress-tensor. 

\medskip

Pulling the questions together:

\begin{itemize}
\item[$\bullet$] Is the QFT content represented by conformally-invariant theory near $\mathcal{I}$?
\item[$\bullet$]Given the presence of $\Lambda$, does the right QFT near $\scri$  have the de Sitter group replacing the Poincar\'e group? (Recall that the de Sitter group {\it{does not}} have a mass Casimir operator.)

 \item[$\bullet$] Can one implement mass fade-out? (Massless charged particles are usually regarded as problematical.)

\item[$\bullet$] Inflation requires a period of exponential expansion just after the big bang, while CCC offers a period of exponential expansion just before the big bang. Can one make sense of the statement \emph{inflation happens before the Bang, in the previous aeon}? And if so, can one then derive the familiar spectrum of density perturbations?

\end{itemize}
This section is all questions and no answers!

\section{Magnetic fields through $\scri$?}
Everywhere astronomers look there are magnetic fields: in the solar system, in stars, in galaxies, in clusters of galaxies and between clusters of galaxies. Given the right initial magnetic fields (i.e. post big bang), there are calculations in the literature, typically based on dynamo mechanisms, to amplify the initial magnetic fields and get agreement with the presently observed values but there is no consensus about the origin of the initial values. Since electromagnetism has good conformal properties, can these initial values be understood as coming through from the previous aeon?

\medskip

As a simple model, consider a flat FLRW metric for $\hg$:
\[\hat{g}=d\hat{t}^2-(\hat{a}(\hat{t}))^2(dx^2+dy^2+dz^2)\]
then as we saw in the Introduction, $\cg$ is also spatially flat FLRW. Now consider the 2-form
\[\hat{B}=dx\wedge dy.\]
This is closed and co-closed, so it's a source-free Maxwell field, and it's orthogonal to the Hubble flow $\partial_{\hat{t}}$ so it's a (very simple) magnetic field. Clearly it extends through $\scri$ with the conventions $\check{B}=\hat{B}$ (which of course works more generally). Note we can confuse ourselves by using orthonormal bases (e.g. ${\bf{\hat{e}}}_1=\hat{a}dx$ etc in $\hM$ and corresponding quantities in $\cM$) when, on the two sides,
\[\hat{B}=(\hat{a})^{-2}{\bf \hat{e}}_1\wedge{\bf \hat{e}}_2,\;\;\check{B}=(\check{a})^{-2}{\bf \check{e}}_1\wedge{\bf \check{e}}_2\]
so $|\hat{B}|$ goes to zero at $\mathcal{I}$ and $|\check{B}|$ diverges there...

\medskip

Does this seamless passage through $\mathcal{I}$ work with more realistic fields (e.g. ones with currents as sources)? Also, when the magnetic field emerges on the hot side (the present aeon) is it washed out by astrophysical processes in its physical surroundings? (There is a conformally-invariant form of MHD , see \cite{sb} or an account in \cite{t3}, which might help answer this.)

\section{Circles in the sky and Hawking points}
There are two sets of observations, \cite{gp1,gp2} and \cite{amn,amnp}, claiming observational support for CCC by finding traces in the CMB maps of events in the previous aeon. The idea for \emph{circles in the sky} is that a merger of supermassive black holes late in the previous aeon would send out a burst of gravitational radiation approximately confined between expanding concentric spheres; this would pass through $\scri$ and register on the last scattering surface; our past light cone would intersect this spherical annulus in an annulus on our sky; thus one should seek annular temperature anomalies. Furthermore, if the events are due to a series of smaller black holes merging with a single very large black hole, one would expect to see the annuli grouped into sets of concentric rings. The two collaborations have different methodologies and different statistical analyses, but they claim rather similar (and positive) findings.

The idea of \emph{Hawking points} arises from consideration of the Hawking evaporation of large black holes in the previous aeon.  As we noted in the Introduction, from a time about ten times the current age of the universe, the whole infinite future is very close to $\scri$ in conformal time. Thus the process of evaporating a black hole of say $10^{10}$ solar masses, which would take $10^{100}$ years in proper time\footnote{$2\times 10^{97}$ years according to Wikipedia.}, happens almost at $\scri$ and all the energy from this radiation is dumped into a small sphere on $\scri$. It will spread out in $\cM$ but register on the last scattering surface only in a ball of radius about 1.6 degrees on our sky\footnote{the ball is small and so subtends at us an angle in radians of approximately $\sqrt{2}\tau_{LS}(\tau_0-\tau_{LS})^{-1}$ where $\tau_{LS}$ and $\tau_0$ are conformal time at last scattering and now, and these are given in \cite{t1}.}. These are what Penrose and collaborators \cite{amnp} call Hawking points and what they claim to have detected in \emph{the same places} on the sky in the Planck data and the WMAP data.

All these observations continue to be controversial and the debate is almost always about statistical significance, since the occurence of the circles and points is usually agreed on by all parties. In the context of this article the questions that arise are about the astrophysics of it all: how energetic would a black hole merger in the previous aeon need to be in order to produce the observed effect in the CMB? They need to be bright enough to be as detected but not so bright as to be visible in the day-time! Also, can one use the observations to construct a detailed picture of the black-hole population in the previous aeon -- how many, how big and how distributed?

The distribution of Hawking points on the sky claimed in \cite{amnp} is strikingly inhomogeneous, as is the distribution of centres of four or more concentric rings in \cite{gp1}. Nonetheless this inhomogeneity is not associated with a disturbance in the CMB homogeneity so one can calculate a consistency check: assume the FLRW model and assume that a supermassive black hole of mass say $M$ dumps all its mass effectively into a point $p$ (or at least, a small region around $p$) on $\scri$, then this mass $M$ will be confined to the ball $B$ of coordinate radius $r=\tau_{LS}$ on the last scattering surface which is the intersection of $J^+(p)$, the future of $p$, with the last scattering surface, which is the 3-surface $\tau=\tau_{LS}$. Knowing the density $\rho$ of the present aeon we can calculate the mass in $B$ due to the (averaged out FLRW) matter. In \cite{t3} I obtained the answer $10^{18}$ solar masses for this but it depends on the (observed) parameters entering $\rho$; now provided $M$ is less than $10^{13}$ solar masses then the perturbation in $\rho$ is less than one in $10^{-5}$ which is what one needs. It is the case that the largest known black holes in the present aeon have mass about $10^{10}$ solar masses but there don't seem to be reasons in principle why an earlier aeon shouldn't have much bigger ones. This leads us into the next section.

\section{Can one construct aeon-to-aeon maps?}
The last two sections have been concerned with how massless fields come through from one aeon to the next. Both these notions fall under the rubric of \emph{aeon-to-aeon} or A2A maps:  can one give an algorithm for the matter density distribution, magnetic field strength and other quantities of interest in the present aeon from knowledge of corresponding quantities in the previous aeon? This would be an A2A map, and it would rely on more-or-less conventional relativistic cosmology and relativistic astrophysics applied to the (correct) equations of CCC.

At first sight, CCC does not explain the approximate homogeneity and isotropy of the universe. As noted at the end of the last section, if there was a super-duper-massive black hole in the previous aeon it would certainly produce an inhomogeneity in the matter density in the present aeon (this was pointed out to me in email correspondence with Sabine Hossenfelder and Tim Palmer, which I gratefully acknowledge). The calculation mentioned above suggests that a black hole of $10^{18}$ solar masses in the previous aeon could give $\delta\rho/\rho\sim 1$ at last scattering in the present aeon. However the period between $\scri$ and the last scattering surface gives enough time to smooth out some lesser inhomogeneities on $\scri$, which are already reduced by the long (infinite in proper time) period of exponential expansion in the previous aeon. For this reason the rather striking inhomogeneity in the distribution of Hawking points on the sky found in \cite{amnp} does not lead to a corresponding inhomogeneity in the matter density measured in the CMB.

Once one has the A2A map one can think about iterating it and ask whether it gives convergence to an almost-homogeneous and almost-isotropic universe? (or indeed to something else!) Since the example of the radiation-filled FLRW universe seems, by the calculation in the Introduction to be a fixed-point of the simplest A2A map one could speculate that convergence to something near this was a good bet. However this certainly \emph{is} speculation, at least for now.

\end{document}